\documentclass[useAMS,usegraphicx,usenatbib]{mn2e}
 
\title[GHZ with radial flows of gas]{ The    galactic habitable zone of  the Milky Way and M31  from chemical evolution models  with gas radial flows}

\author[ Spitoni, Matteucci, Sozzetti]{E.~Spitoni,$^1$\thanks{E-mail: spitoni@oats.inaf.it}
  F.~Matteucci$^{1, 2, 3}$, A. Sozzetti $^{4}$\\
  $^1$ Dipartimento di Fisica, Sezione di Astronomia, Universit\`a di Trieste, via G.B. Tiepolo 11, I-34131, Trieste, Italy \\
  $^2$ I.N.A.F. Osservatorio
  Astronomico di Trieste, via G.B. Tiepolo 11, I-34131, Trieste,
  Italy\\
 $^3$ I.N.F.N. Sezione di Trieste, via Valerio 2, 34134 Trieste, Italy\\
$^4$ I.N.A.F. - Osservatorio Astrofisico di Torino, Via Osservatorio 20, I-10025 Pino Torinese, Italy
}
\usepackage[percent]{overpic}

\begin{document}
\date{Accepted . ; in original form xxxx}

\pagerange{\pageref{firstpage}--\pageref{lastpage}} \pubyear{xxxx}

\maketitle

\label{firstpage}

\begin{abstract}
The galactic habitable zone is defined as the region with sufficient
abundance of heavy elements to form planetary systems in which
Earth-like planets could be born and might be capable of sustaining
life, after surviving to close supernova explosion events. Galactic
chemical evolution models can be useful for studying the galactic
habitable zones in different systems. We apply detailed chemical
evolution models including radial gas flows to study the galactic
habitable zones in our Galaxy and M31. We compare the results to the
relative galactic habitable zones found with ``classical'' (independent ring) models, where
no gas inflows were included. For both  the Milky Way and Andromeda,
the main effect of the gas radial inflows is to enhance the number of
stars hosting a habitable planet with respect to the ``classical'' model
results, in the region of maximum probability for this occurrence,
relative to the classical model results. These results are obtained by
taking into account the supernova destruction processes.  In
particular, we find that in the Milky Way the maximum number of stars
hosting habitable planets is at 8 kpc from the Galactic
center, and the model with radial flows predicts a number which is
38\% larger than what predicted by the classical model. For Andromeda we find that the
maximum number of stars with habitable planets is at 16 kpc from the
center and that in the case of radial flows this number is larger by
10 \% relative to the stars predicted by the classical model.

\end{abstract}


\section{Introduction}

The Circumstellar Habitable Zone (CHZ) has generally been defined to
be that region around a star where liquid water can exist on the
surface of a terrestrial (i.e., Earth-like) planet for an extended
period of time (Huang 1959, Shklovsky \& Sagan 1966, Hart 1979).
Kasting et al. (1993) presented the first one-dimensional climate
model for the calculation of the width of the CHZ around the Sun and
other main sequence stars.  Later on, several authors improved that
model: Underwood et al. (2003) computed the evolution of the CHZ
during the  evolution of the host star. Moreover, Selsis et
al. (2007) considered the case of low mass stars, and Tarter et
al. (2007) asserted that M dwarf stars can host planets in which the
origin and evolution of life can occur.
Recently, Vladilo et
al. (2013) applied a one-dimensional energy balance model to
investigate the surface habitability of planets with an Earth-like
atmospheric composition but different levels of surface pressure.

  However what is most relevant to our paper is that it exists a
  well-established correlation between metallicity of the stars and
  the presence of giant planets: the host stars are more metallic than
  a normal sample ones (Gonzalez, 1997; Gonzalez et al., 2001; Santos
  et al., 2001, 2004; Fischer \& Valenti, 2005; Udry et al., 2006;
  Udry \& Santos, 2007). In particular, Fisher \& Valenti (2005) and
  Sousa at al. (2011) presented the probabilities of the formation of
  giant planets as a function of [Fe/H] values of the host star.  In
  Sozzetti et al. 2009, Mortier et al. (2013a,b) these probabilities
  are reported for different samples of stars.

 The galactic chemical evolution can substantially influence the
  creation of habitable planets. In fact, the model of Johnson \& Li
  (2012) showed that  the first Earth-like planets likely formed
  from circumstellar disks with metallicities $Z \geq 0.1
  Z_{\odot}$. Moreover, Buchhave et al. (2012), analyzing the mission
  Kepler, found that the frequencies of the planets with earth-like
  sizes are almost independent of the metallicity, at least up to
  [Fe/H] values $\sim$ 0.6 dex. This is it was confirmed by Sousa et
  al. (2011) from the radial velocity data.  Data from Kepler and from
  surveys of radial velocities from earth show that the frequencies of
  planets with masses and radii not so different from the Earth, and
  with habitable conditions are high: $\sim 20 \%$ for stars like the
  Sun (Petigura et al. 2013), between the 15 \% (Dressing \&
  Charbonneau 2013) and 50 \% for M dwarf stars (Bonfils et
  al. 2013).

Habitability on a larger scale was considered for the first time by
Gonzalez et al. ( 2001), who introduced the concept of the galactic
habitable zone (GHZ).

The GHZ is defined as the region with
sufficient abundance of heavy elements to form planetary systems in
which Earth- like planets could be found and might be capable of
sustaining life. Therefore, a minimum metallicity is needed for
planetary formation, which would include the formation of a planet
with Earth-like characteristics (Gonzalez et al. 2001, Lineweaver
2001, Prantzos 2008).  

 Gonzalez et al. (2001) estimated, very approximately, that a
metallicity at least half that of the Sun is required to build a
habitable terrestrial planet and the mass of a terrestrial planet has
important consequences for interior heat loss, volatile inventory, and
loss of atmosphere.

On the other hand, various physical processes may favor the destruction
of life on planets. For instance, the risk of a supernova (SN)
explosion sufficiently close represents a serious risk for the life
(Lineweaver et al. 2004, Prantzos 2008, Carigi et al. 2013).

Lineweaver et al. (2004), following the prescription of Lineweaver et
al. (2001) for the probability of earth-like planet formation,
discussed the GHZ of our Galaxy.  They modeled the evolution of the
Milky Way in order to trace the distribution in space and time of four
prerequisites for complex life: the presence of a host star, enough
heavy elements to form terrestrial planets, sufficient time for
biological evolution, and an environment free of life-extinguishing
supernovae. They identified the GHZ as an
annular region between 7 and 9 kpc from the Galactic center
that widens with time.

Prantzos (2008) discussed the GHZ for the Milky Way. The role of
metallicity of the protostellar nebula in the formation and presence
of Earth-like planets around solar-mass stars was treated with a new
formulation, and a new probability of having Earths as a function of
[Fe/H] was introduced.

In particular, Prantzos (2008) criticized the modeling of GHZ  based on the idea of destroying life permanently by SN explosions.

Recently, Carigi et al. (2013) presented a model for the GHZ of
M31. They found that the most probable GHZ is located between 3 and 7
kpc from the center of M31 for planets with ages between 6 and 7
Gyr. However, the highest number of stars with habitable planets was
found to be located in a ring between 12 and 14 kpc with a mean age of
7 Gyr. 11\% and 6.5\% of all the formed stars in M31 may have planets
capable of hosting basic and complex life, respectively. However,
Carigi at al. (2013) results are obtained using a simple chemical
evolution model built with the instantaneous  recycling approximation
which does not allow to follow the evolution of Fe, and where no
inflows of gas are taken into account.

In this work we investigate for the first time the effects of radial
flows of gas on the galactic habitable zone for both our Galaxy 
and M31, using detailed chemical evolution models in which is relaxed
the instantaneous recycle approximation, and the core collapse and
Type Ia SN rates are computed in detail.  For the GHZ calculations
  we use the Prantzos (2008) probability to have life around a star
  and the Carigi et al (2013) SN destruction effect prescriptions.

 In this work we do not take into account the possibility of
  stellar migration (Minchev et al. 2013, and Kybrik at al. 2013) and
  this will be considered in a forthcoming paper.

The paper is organized as follows: in Sect. 2 we present our galactic
habitable zone model, in Sect. 3 we describe our ``classical'' chemical
evolution models for our Galaxy and M31, in Sects. 4 the reference
chemical evolution models in presence of radial flows are
shown. The results of the galactic habitable zones for the
``classical'' models are presented in Section. 5, whereas the ones in
presence of radial gas flows in Section 6.  Finally, our
  conclusions are summarized in Section 7.

\section{The galactic habitable zone model}

\begin{figure}
	  \centering   
    \includegraphics[scale=0.35]{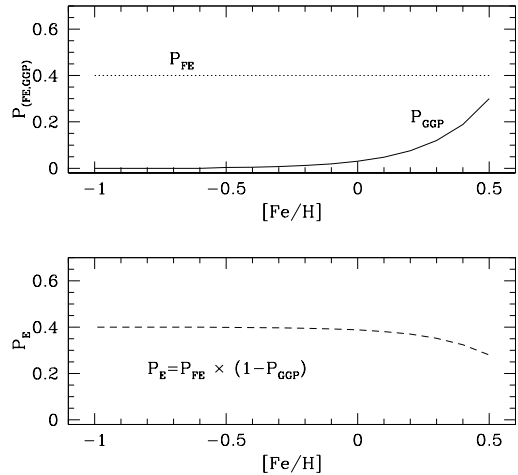} 
  \caption{{\it Upper panel}: The probabilities of forming Earths
    $P_{FE}$, and forming gas giant planets $P_{GGP}$ are
    reported. {\it Lower panel}: Following Prantzos (2008) the
    probability of having Earths $P_{E}$, obtained from the two
    previous ones.}
		\label{prob}
\end{figure}

 Following the assumptions of Prantzos et al. (2008) we define
 $P_{FE}$ as the probability of forming Earth-like planets which is a
 function of the Fe abundance: the $P_{FE}$ ($FE$ means forming
 earths)value is 0.4 for [Fe/H] $\geq$ -1 dex, otherwise $P_{FE}=0$
 for smaller values of [Fe/H].  This assumption is totally in
   agreement with the recent results about the observed  metallicities of
   Earth-like planets  presented in the Introduction (Petigura et al. 2013, Dressing \& Charbonneau 2013, Bonfilis et al. 2013).

\begin{figure*}
	  \centering   
    \includegraphics[scale=0.3]{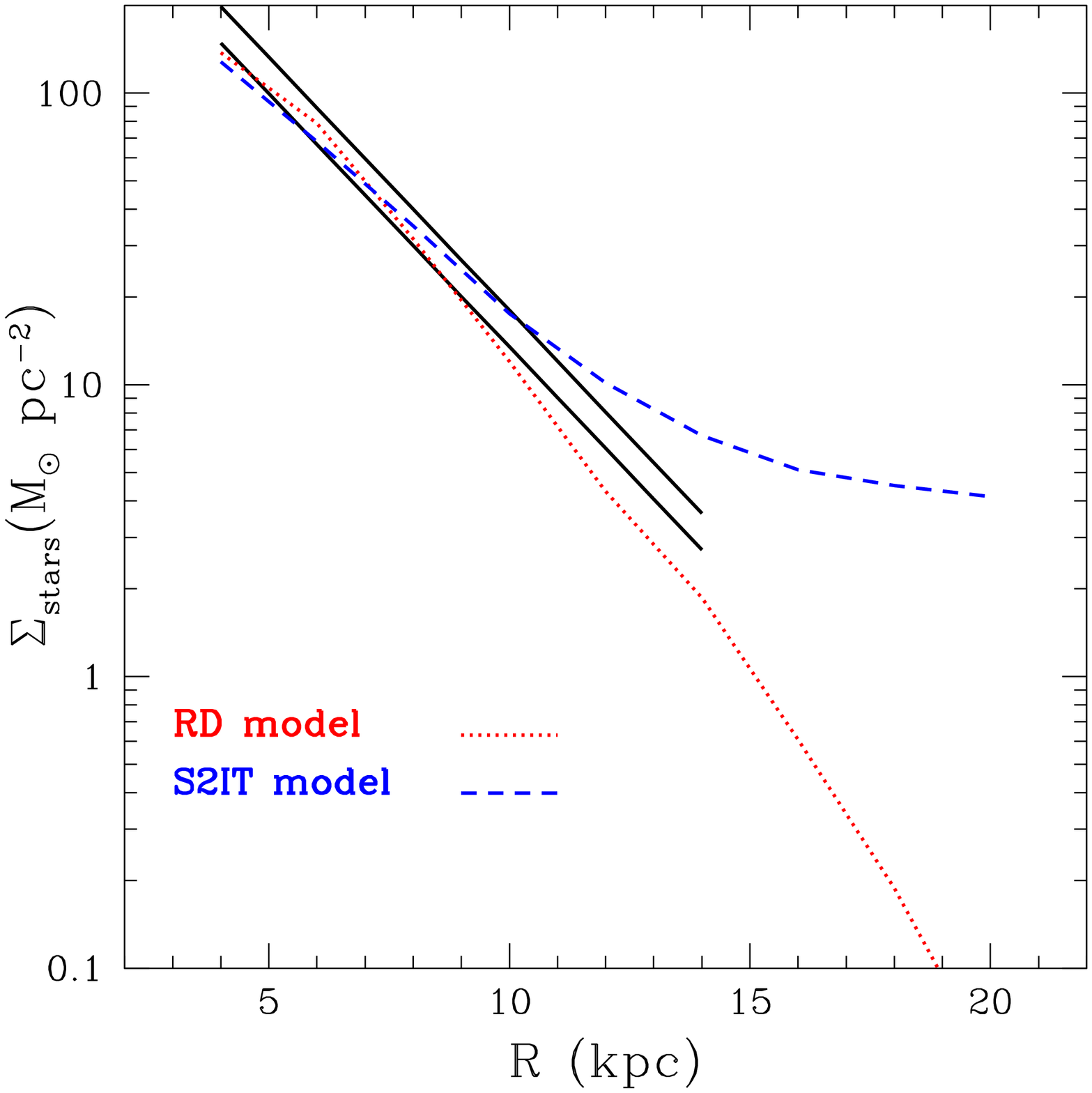} 
     \includegraphics[scale=0.3]{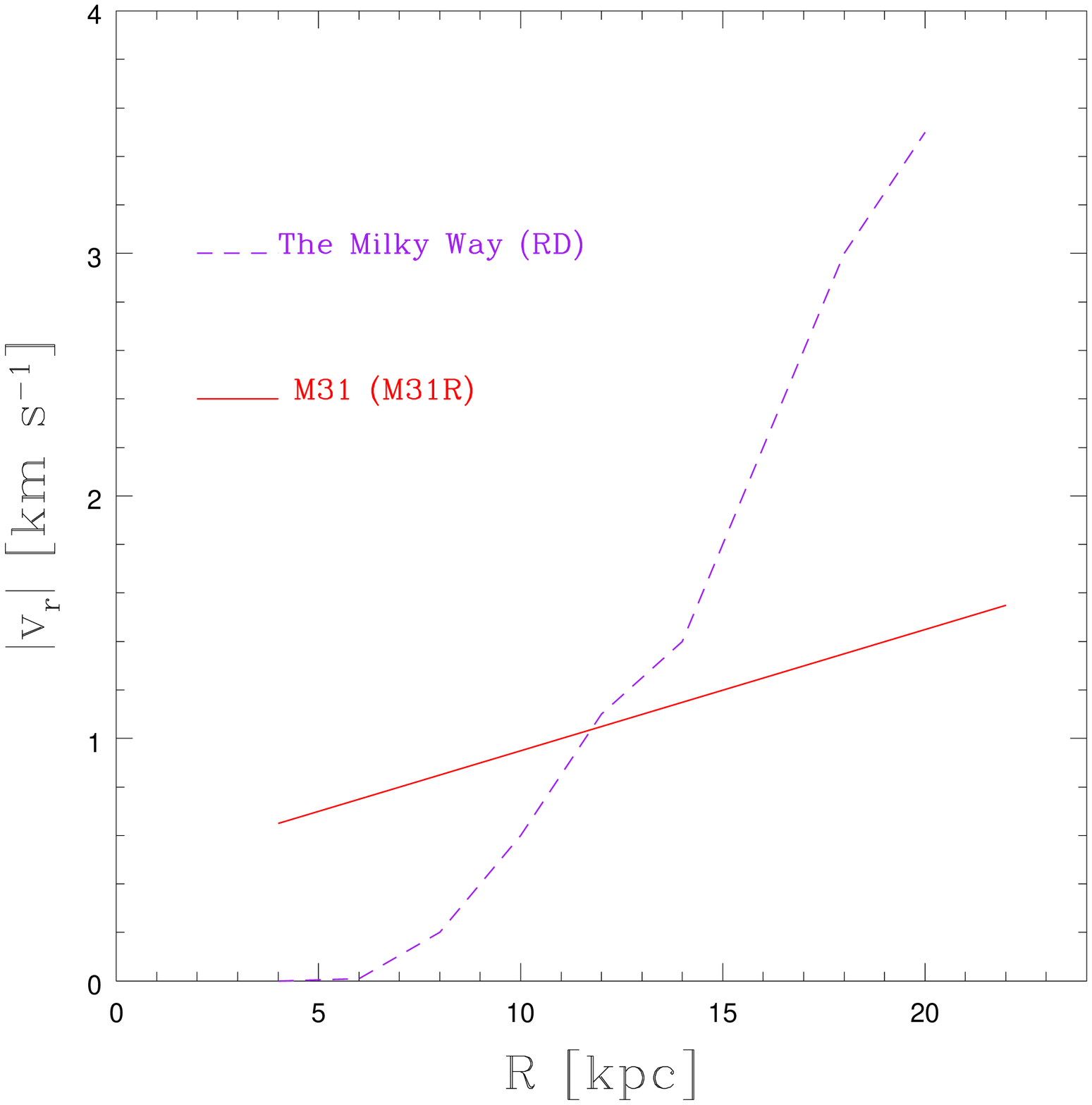} 
    \caption{{\it Left panel}: Observed radial stellar density profile
      for the Milky Way is drawn with solid black lines, the
      prediction of our ``classical'' model S2IT is shown with the blue
      dashed line, whereas the model with radial gas flows with the
      red dotted line. {\it Right panel}: Velocity pattern for the
      radial inflows of gas for The Milky Way model RD (dashed line)
      and for the M31 model M31B one (solid line). }
		\label{vel}
\end{figure*}

Fischer \&  Valenti (2005) studied  the probability of formation  of a
gaseous giant planet which is a function of metallicity. In particular
they  found the  following  relation for  FGK  type stars  and in  the
metallicity range -0.5 $<$ [Fe/H] $<$ 0.5:

\begin{equation}
 P_{GGP}\left(\mbox{[Fe/H}]\right)= 0.03 \times 10^{2.0 \mbox{[Fe/H]}},
\label{valenti}
\end{equation}
where $GGP$ stands for gas giant planets.  The $P_{FE}$, and $P_{GGP}$
probabilities vs [Fe/H] used in this work are reported in the upper
panel of Fig. \ref{prob}.
 
 Prantzos (2008) identified $P_{GGP}$ (eq. \ref{valenti}) from Fisher \& Valenti
  (2005) with the probability of the formation of hot jupiters, althought this assumption could be questionable. In fact this probability should be related
 to the formation of giant gas planets in general.
 However, Carigi et al. (2013) computed the GHZ for M31 testing different
 probabilities of terrestrial planet formation taken by Lineweaver et
 al. (2004), Prantzos (2008) and the one from The Extrasolar Planets
 Encyclopaedia on March 2013. As it can be seen from their Fig.8, the
 choice of different probabilities does not modify in a substantial
 way the GHZ.

In the lower panel of Fig. \ref{prob} the probability of having stars
with Earth like planets but not gas giant planets which destroy them, is
reported as $P_E$. This quantity is simply given by:

\begin{equation}
P_E=P_{FE} \times (1-P_{GGP}).
\end{equation}

We define $P_{GHZ}(R,t)$ as the fraction of all stars
having Earths (but no gas giant planets) which survived supernova
explosions as a function of the galactic radius and time:

\begin{equation}
P_{GHZ}(R,t)= \frac{\int_0^t SFR(R,t') P_E (R,t') P_{SN}(R,t') dt'}{\int_0^t SFR(R,t')dt'}
\label{GHZ}
\end{equation}

This quantity must be interpreted as the relative
probability to have complex life around one star at a given position,
as suggested by Prantzos (2008).

In eq. (\ref{GHZ}) $SFR(R,t')$ is the star formation rate (SFR) at the
time $t'$ and galactocentric distance $R$, $P_{SN}(R,t')$ is the probability of
surviving supernova explosion.

For this quantity we refer to the work of Carigi et al. (2013).  The
authors explored different cases for the life annihilation on formed
planets by SN explosions. Among those they assumed that the SN
destruction is effective if the SN rate at any time and at any radius
has been higher than the average SN rate in the solar neighborhood
during the last 4.5 Gyr of the Milky Way's life (we call it
$<RSN_{SV}>$).

 Throughout our paper we refer to this condition as ``case 1)''.

  Because of the uncertainties about the real effects of SNe on life
  destruction, we also tested a case in which the annihilation is
  effective if the SN rate is higher than $2 \times <RSN_{SV}>$, and
  we call it ``case 2)''.  This condition is almost the same as that
  used by Carigi et al. (2013) to describe their best models. They
  imposed that, since life on Earth has proven to be highly resistant,
  there is no life if the rate of SN during the last 4.5 Gyr of the
  planet life is higher than twice the actual SN rate averaged over
  the last 4.5 Gyr ($2 \times <RSN_{SV}>$).

We associate the following probabilities $P_{SN}(R,t)$ at the cases 1)
and 2), respectively:

\begin{itemize}
\item Case 1): if the SN rate is larger than $<RSN_{SV}>$ then
  $P_{SN}(R,t)$ = 0 else $P_{SN}(R,t)$ = 1;

\item Case 2): if the SN rate is larger than $2 \times <RSN_{SV}>$
  then $P_{SN}(R,t)$ = 0 else $P_{SN}(R,t)$ = 1.
\end{itemize}

\begin{table*}
\caption{Chemical evolution models for the Milky Way}
\scriptsize 
\label{TMW}
\begin{center}
\begin{tabular}{c|cccccccc}
  \hline
\\
 Models &Infall type &$\tau_d$& $\tau_H$&$\nu$&Threshold & $\sigma_H(R)$&Radial inflow  \\
\\
&  &[Gyr]& [Gyr]&[Gyr$^{-1}$]& [M$_{\odot}$pc$^{-2}$]&[M$_{\odot}$pc$^{-2}$] &[km s$^{-1}$]\\  
  
\hline

S2IT & 2 infall & 1.033 R[kpc]-1.27  &0.8  & 1 &7 (thin disc) & 17& /\\
Spitoni \& Matteucci (2011)&&&&& 4 (halo-thick disk)\\
\hline

RD&  2 infall & 1.033 R[kpc]-1.27    & 0.8&1  &/& 17 if $R \leq 8$ kpc& velocity pattern III\\
Mott et al. (2013)&&&&&& 0.01 if $R \geq 10$ kpc&Mott et al. (2013)\\
 \hline
\end{tabular}
\end{center}
\end{table*}

\begin{table*}
\caption{Chemical evolution models for M31}
\scriptsize 
\label{TM31}
\begin{center}
\begin{tabular}{c|cccccc}
  \hline
\\
 Models & Infall type&$\tau$& $\nu$&Threshold &Radial inflow  \\
\\
& & [Gyr]& [Gyr$^{-1}$]& [M$_{\odot}$pc$^{-2}$]&  [km s$^{-1}$]\\  
  
\hline

M31B &1 infall  &0.62 R[kpc] +1.62   & 24/(R[kpc])-1.5  &5 & /\\
Spitoni et al. (2013)&&&&\\
\hline

M31R&1 infall&   0.62 R[kpc] +1.62  & 2  &/& $v_R$ = 0.05 R[kpc] + 0.45 \\
Spitoni et al. (2013)&&&&\\
 \hline
\end{tabular}
\end{center}
\end{table*}

For $<RSN_{SV}>$ we adopt the value of 0.01356 Gyr$^{-1}$ pc$^{-2}$
using the results of the S2IT model of Spitoni \& Matteucci (2011).
Some details of this model will be provided in Section 3.1. Here, we
just recall that in this model the Galaxy is assumed to have formed by
means of two main infall episodes: the first formed the halo and the
thick disk, and the second the thin disk.

 In Carigi et al. (2013) work it was
considered for $<RSN_{SV}>$ a value of 0.2 Gyr$^{-1}$ pc$^{-2}$, we
believe it is a just a typo and all their results are obtained with
the correct value for $<RSN_{SV}>$.

We consider also the case where the effects of SN explosions are not
taken into account, and with this assumption eq. (\ref{GHZ}) simply
becomes:
\begin{equation}
P_{GHZ}(R,t)= \frac{\int_0^t SFR(R,t') P_E (R,t')  dt'}{\int_0^t SFR(R,t')dt'}
\label{GHZ_NOSN}
\end{equation}

Detailed chemical evolution models can be an useful tool to
estimate the GHZ for different galactic systems. In the next two
Sections we present the models  for our Galaxy and for M31. We will call
``classical'' the models in which no radial inflow of gas was considered.

 Finally, we define the total number of stars formed at a certain time $t$ and galactocentric distance $R$  hosting Earth-like planet with life $N_{\star \, life}(R,t)$, as:

\begin{equation}
N_{\star \, life}(R,t)=P_{GHZ}(R,t) \times N_{\star tot}(R,t), 
\end{equation} 
where  $N_{\star tot}(R,t)$ is the total number of stars created up to  time $t$ at the galactocentric distance $R$.

\section{The  ``classical'' chemical evolution models}

In this section we present the best ``classical'' chemical evolution
models we will use in this work.  For the Milky Way our reference
classical model is the ``S2IT'' of Spitoni \& Matteucci (2011), whereas
for M31 we refer to the model ``M31B'' we proposed in Spitoni et
al. (2013).
 
\begin{figure}
	  \centering   
    \includegraphics[scale=0.35]{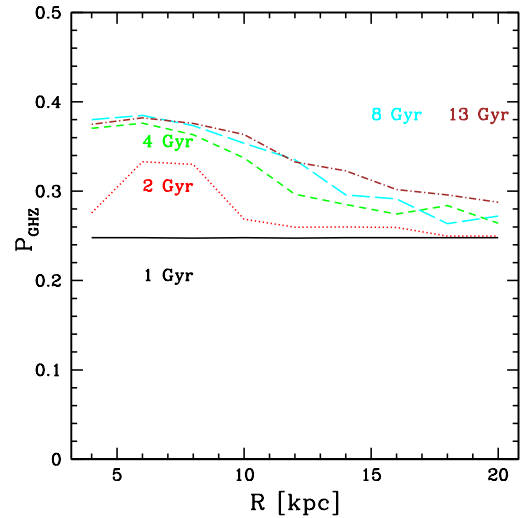} 
    \caption{The probability $P_{GHZ}(R,t)$ as a function of the galactocentric distance of having Earths  where the effects of SN explosions are not considered for the classical model S2IT at 1, 2, 4, 8, 13 Gyr.}
		\label{S2IT_NOSN_t}
\end{figure}

\begin{figure}
	  \centering   
    \includegraphics[scale=0.35]{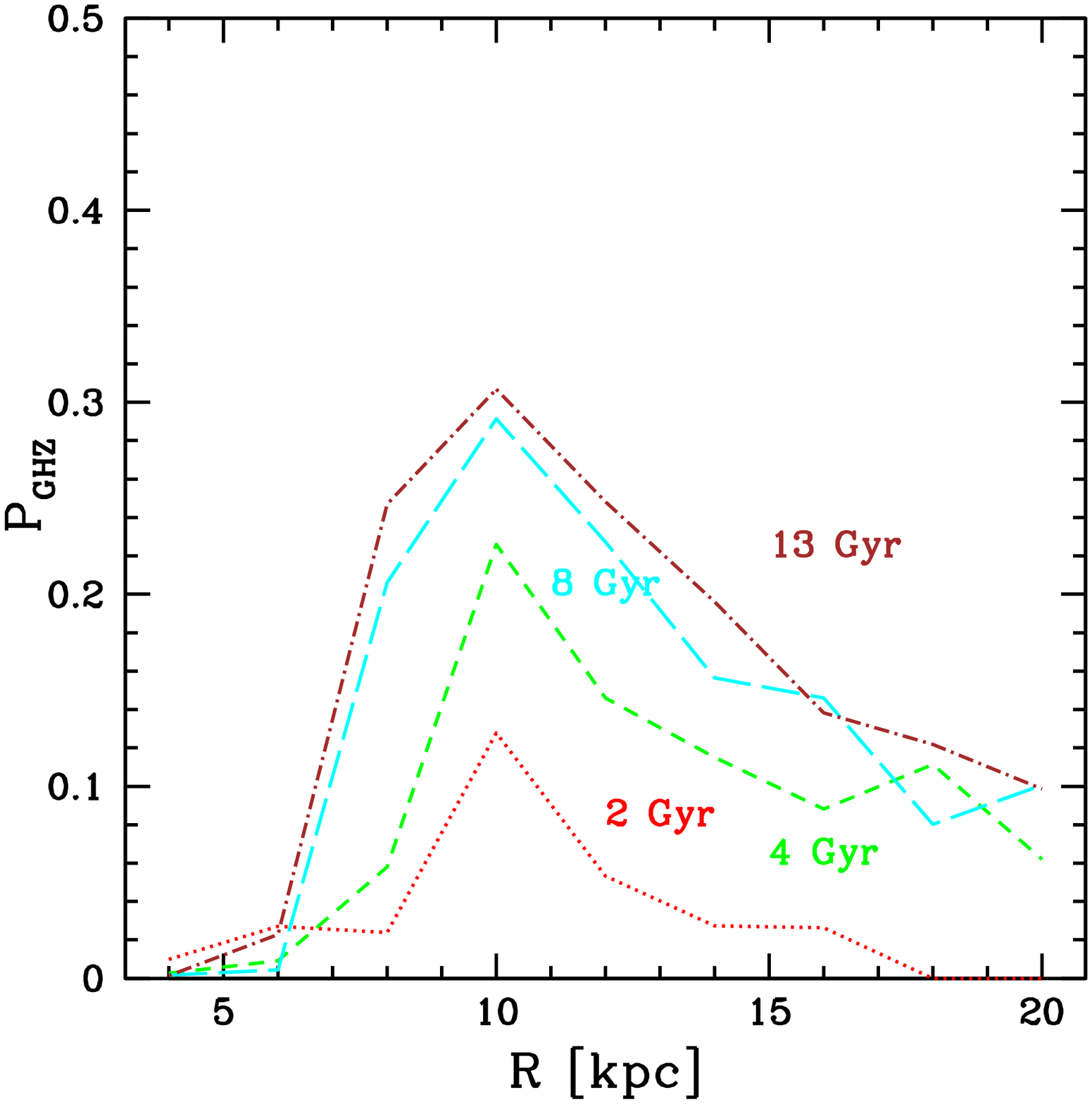} 
    \caption{The probability $P_{GHZ}(R,t)$ as a function of the
      galactocentric distance of having Earths where the case 2) of SN
      explosions are considered for the classical model S2IT at 1, 2, 4,
      8, 13 Gyr.}
		\label{S2IT_SN2_t}
\end{figure} 

\begin{figure}
	  \centering   
    \includegraphics[scale=0.35]{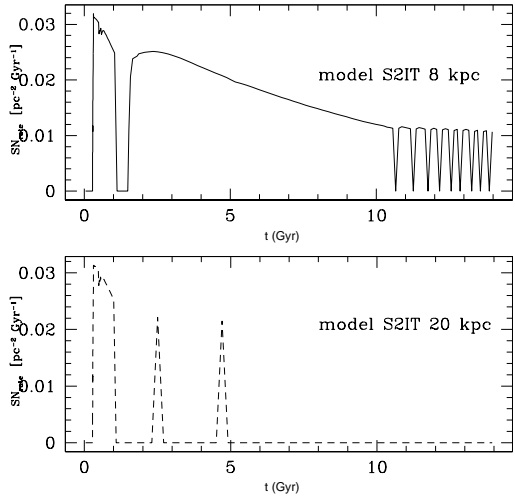} 
    \caption{A comparison between the SN rate histories at 8  and 20 kpc for the model S2IT.}
		\label{SN_S2IT_8_20}
\end{figure}

\subsection{The Milky Way (S2IT model)}
To follow the chemical evolution of the Milky Way without radial flows
of gas, we adopt the model S2IT of Spitoni \& Matteucci (2011) which
is an updated version of the two infall model Chiappini et al. (1997)
model.   This model assumes that the halo-thick disk forms out of an
  infall episode independent of that which formed the disk. In
  particular, the assumed infall law is

 \begin{equation}
A(r,t)= a(r) e^{-t/ \tau_{H}}+ b(r) e^{-(t-t_{max})/ \tau_{D}(r)},
\end{equation}
where $\tau_{H}$  is  the typical timescale for the formation of the halo and thick disk is
0.8 Gyr, while $t_{max}$ = 1 Gyr is the time for the
maximum infall on the thin disk.
The coefficients $a(r)$ and $b(r)$ are obtained by imposing a
fit to the observed current total surface mass density in the thin
disk as a function of galactocentric distance given by:
\begin{equation}
\Sigma(r) = \Sigma_0 e^{-R/R_D},
\end{equation}
where $\Sigma_0$ = 531 M pc$^{-2}$ is the central total
surface mass density and $R_D$ = 3.5 kpc is the scale length.

 Moreover, the
formation timescale of the thin disk is assumed to be a function of
the Galactocentric distance, leading to an inside-out scenario for the
Galaxy disk build-up. The Galactic thin disk is approximated by
several independent rings, 2 kpc wide, without exchange of matter
between them. 

\begin{figure} 
	  \centering   
    \includegraphics[scale=0.55, angle=-90]{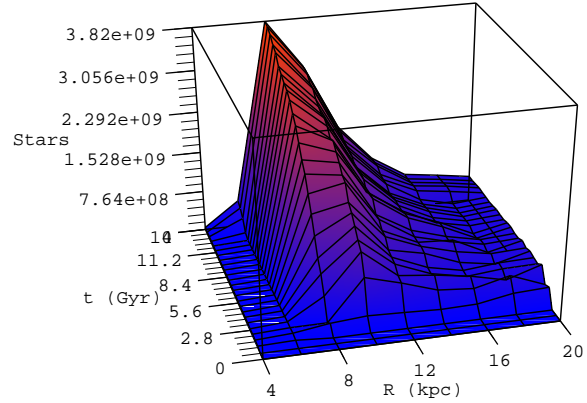} 
    \caption{The total number of stars  having Earths ($N_{\star life}$) as a function of the Galactocentric distance and the Galactic time 
      where the case 2) of SN explosions are considered for the classical
      model S2IT. The ($N_{\star life}$) values  are computed within concentric rings, 2 kpc wide.}
		\label{map_S2IT_SN2}
\end{figure}

A threshold gas density of 7 M$_{\odot}$ pc$^{-2}$ in the SF process
(Kennicutt 1989, 1998; Martin \& Kennicutt 2001; Schaye 2004) is also
adopted for the disk. The halo has a constant surface mass density as
a function of the galactocentric distance at the present time equal to
17 M$_{\odot}$ pc$^{-2}$ and a threshold for the star formation in
the halo phase of 4 M$_{\odot}$ pc$^{-2}$, as assumed for the model B
of Chiappini et al. (2001).

 The assumed IMF is the one of Scalo (1986), which is assumed  constant in
  time and space.  The adopted law for the SFR is
  a Schmidt (1958) like one:
\begin{equation}
\Psi \propto \nu \Sigma^k_{gas}(r,t),
\label{schmidt}
\end{equation}

where $\Sigma_{gas}(r,t)$ is the surface gas density with the exponent
$k$ equal to 1.5 (see Kennicutt 1998; and Chiappini et al. 1997). The
quantity $\nu$ is the efficiency of the star formation process, and it
is constant and fixed to be equal to 1 Gyr $^{-1}$. 
 
 In Table \ref{TMW} the principal characteristics of the S2IT model
 are reported: in the second column the infall type is reported, in
 the third and forth ones the time scale $\tau_d$ of the thin disk
 formation, and the time scale $\tau_H$ of the halo formation are
 drawn. The dependence of $\tau_d$ on the Galactocentric distance $R$,
 as required by the inside-out formation scenario, is expressed in the
 following relation:
\begin{equation}
\tau_d= 1.033 R\mbox{[kpc]}-1.27 \mbox{ [Gyr]}.
\end{equation}

  is in column 5. The adopted threshold in the surface gas density for
  the star formation, and total surface mass density for the halo are
  reported in column 6 and 7, respectively. 

 In the left panel of Fig. \ref{vel} the surface density profile for
 the stars as predicted by the model S2IT is reported. Observational
 data are the ones used by Chiappini et al. (2001).

\begin{figure}
	  \centering   
    \includegraphics[scale=0.35]{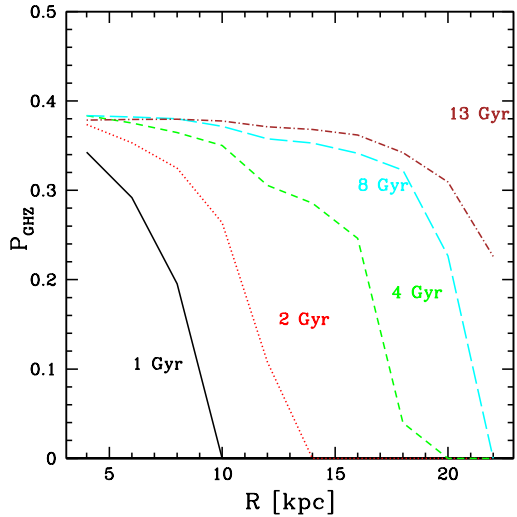} 
    \caption{The probability $P_{GHZ}(R,t)$ as a function of the
      galactocentric distance of having Earths where the effects of SN
      explosions are not considered for the classical model M31B at 1, 2,
      4, 8, 13 Gyr.}
		\label{M31B_NOSN_t}
\end{figure}

\subsection{M31 (M31B model)}
 
To reproduce the chemical evolution of M31, we adopt the best model
M31B of Spitoni et al. (2013).  The surface mass density distribution
is assumed to be exponential with the scale-length radius $R_D=5.4$
kpc and central surface density $\Sigma_0 = 460 M_{\odot}pc^{-2}$, as
suggested by Geehan et al. (2006). It is a one infall model with
inside-out formation, in other words we consider only the formation of
the disk. The time scale for the infalling gas is a function of the
Galactocentric radius: $\tau(R)=0.62R+1.62$. The disk is divided in
several shells 2 kpc wide as for the Milky Way.

 In order to reproduce the gas distribution they adopted for the model
 M31B the SFR of eq. (\ref{schmidt}) with the following star formation efficiency: $\nu(R) = 24/R - 1.5$, until it reaches a minimum value
 of 0.5 Gyr$^{-1}$ and then is assumed to be constant.  Finally, a
 threshold in gas density for star formation of 5 $M_{\odot}/pc^{2}$
 is considered, as suggested in Braun et al. (2009).  The model
   parameters for the time scale of the infalling gas, for the star
   formation efficiency, and for the threshold are summarized in Table
   \ref{TM31}. The assumed IMF is the one of Kroupa et al. (1993).

 Our reference model of M31 overestimates the present-day SFR
  data. In fact, our ``classical'' model is similar to that of
  Marcon-Uchida et al. (2010). The same model, however, well
  reproduces the oxygen abundance along the M31 disk. We think that
 more  uncertainties are present in the derivation of the star
  formation rate than in the abundances.

\section{The  chemical evolution models with radial flows of gas}

In this section we present the best models in presence of radial gas
flows we will use in this work. For the Milky
Way our reference radial gas flow model is the ``RD'' of Mott et
al. (2013), whereas for M31 we refer to the model
``M31R'' we proposed in Spitoni et al. (2013).

\subsection{The Milky Way (RD model)}

We consider here the best model RD presented by Mott et al. (2013) to
describe the chemical evolution of the Galactic disk in presence of
radial flows. From Table \ref{TMW} we see that this model shows the
same prescriptions of the S2IT model for the inside-out formation and
for the SFR efficiency fixed at 1 Gyr$^{-1}$. On the other hand the RD
model has not a threshold in the SF and a different modeling of the
surface density for the halo was used: the total surface mass density
in the halo $\sigma_H(R)$ becoming very small for distances $\geq$ 10
kpc, a more realistic situation than that in model S2IT.

The radial flow velocity pattern is shown in Fig.  \ref{vel} labeled
as ``pattern III''.  The range of velocities span between 0 and 3.6 km
s$^{-1}$.

We recall here that in the implementation of the radial inflow of gas
in the Milky Way, presented by Mott et al. (2013), only the gas that
resides inside the Galactic disk within the radius of 20 kpc can move
inward by radial inflow, and as a boundary condition we impose that
there is no flow of gas from regions outside the ring centered at 20
kpc.

 In the left panel of Fig. \ref{vel} the surface density profile
  for the stars predicted by the model RD is reported. Observational
  data are the ones used by Chiappini et al. (2001).

\subsection{M31 (M31R model)}

M31R is the best model for M31 in presence of radial flows presented
by Spitoni et al. (2013). It assumes a  constant star formation efficiency,
fixed at the value of 2 Gyr$^{-1}$ and, it does not include a star
formation threshold.

 \begin{figure*}
	  \centering   
    \includegraphics[scale=0.3]{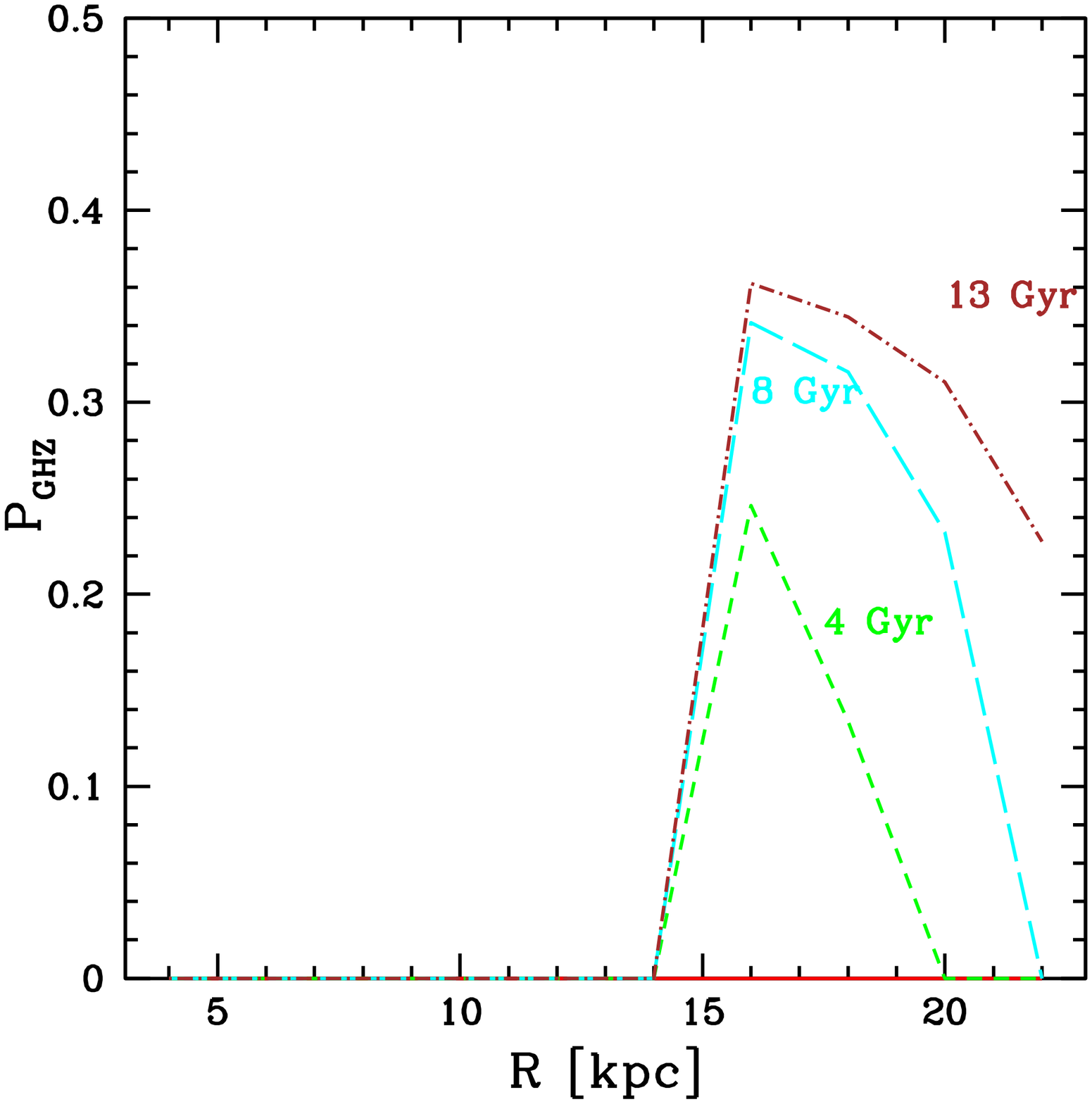} 
     \includegraphics[scale=0.3]{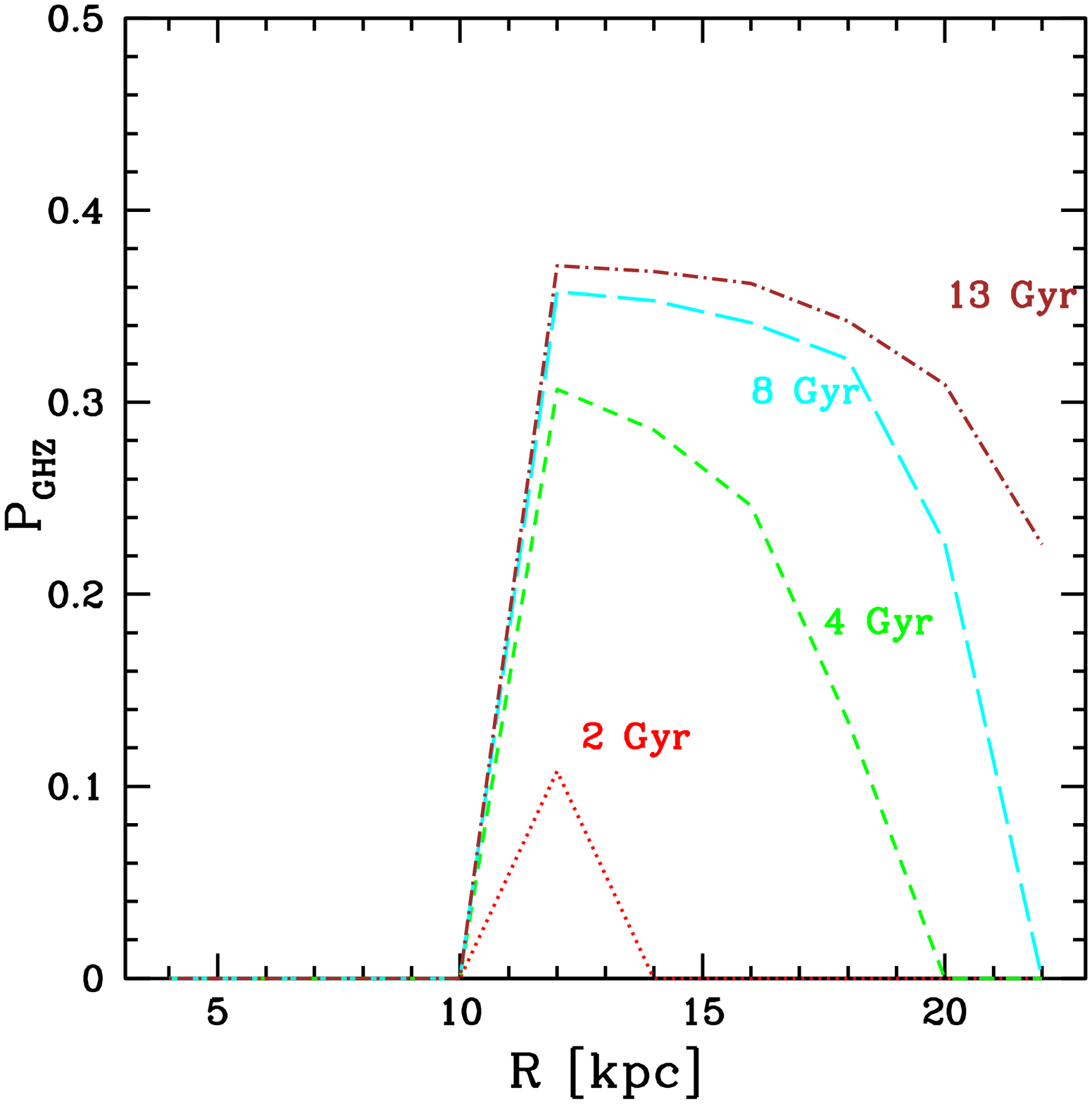} 
    \caption{ The probability $P_{GHZ}(R,t)$ as a
        function of the galactocentric distance of having Earth-like planets
        for the classical model M31B at 1, 2, 4, 8, 13 Gyr including the case 1)  SN explosions effects  in the left panel. The  results for the same model but with   the  case 2) prescription for the SN destruction  are drawn in the right panel.}
		\label{M31B_SN1_t}
\end{figure*}

At  variance with the RD model, where the radial flows  was
applied to a two-infall model, for the M31R model  it was possible to find a
linear velocity pattern as a function of the galactocentric
distance. 

The radial inflow velocity pattern requested to
reproduce the data follows this linear relation:

\begin{equation}
v_R = 0.05 R[\mbox{kpc}] + 0.45\mbox{  }  [\mbox{km s}^{-1}],
\label{linear}
\end{equation}
and spans the range of velocities between 1.55 and 0.65 km s$^{-1}$ as shown in Fig. \ref{vel}.

Therefore in the external regions the velocity inflow for
the Milky Way model is higher than the M31 velocity flows
as shown in Fig. \ref{vel}.  At 20 kpc the ratio between the inflow velocities is $v_{RD}/v_{M31R} \simeq 2.5$. 

The model M31R fits the O abundance gradient in the disk of M31
very well. The other model parameters are reported in Table \ref{TM31}.

We recall here that in the implementation of the radial inflow of gas
in M31 presented by Spitoni et al. (2013), only the gas that resides
inside the Galactic disk within the radius of 22 kpc can move inward
by radial inflow, and as boundary condition we impose that there is no
flow of gas from regions outside the ring centered at 22 kpc, as already discussed for the Milky Way.

 \section{The Classical model GHZ results}

In this section we report our results concerning the GHZ using
``classical'' chemical evolution models.

\subsection{The Milky Way model results}

We start to present the Milky Way results for the models without any
radial flow of gas.  In Fig. \ref{S2IT_NOSN_t} the probability
$P_{GHZ}(R,t)$ for the model S2IT of our Galaxy without the effects of
SN are reported, at 1, 2, 4, 8, 13 Gyr.  We recall that S2IT is a two
infall model. Comparing Table \ref{TMW} with Table \ref{TM31} we
notice another important difference between the S2IT and M31B
models. For the Milky Way the star formation efficiency $\nu$ is
constant and taken equal to 1 Gyr$^{-1}$ whereas for M31 is $\nu$=
24/(R[kpc])-1.5.  This is the reason why our results are different
from the Prantzos (2008) ones.  In fact the reference chemical
evolution model for the Milky Way in Prantzos (2008) is the one
described in Boissier \& Prantzos (1999), where the SFR is $\propto
R^{-1}$.  Hence at variance with the Prantzos (2008) GHZ results, the
probability that a star has a planet with life is high also in the
external regions at the early times.   At 1 Gyr we notice that the
  $P_{GHZ}$ values  are constant along the disk.  The reason
for that resides in the fact that in the first Gyr the SFR is the same
at all radii, since it reflects the SF in the halo (see Fig.1  of Spitoni et al. 2009).

For our Galaxy we just show the results of the SN case 2) model
reported in Fig. \ref{S2IT_SN2_t}.  This is
our best model for the Milky Way considering our SN rate history. With
this model the region with the highest probability that a formed star
can host a Earth-like planet with life is between 8 and 12 kpc, and
the maximum located at 10 kpc.

 The Milky Way outer parts are affected by the SN destruction: in
 fact, the predicted SN rate using Galaxy evolution as time goes by
 can overcome the value fixed by the case 2): $2 \times<RSN_{SV}>$,
 and consequently $P_{GHZ}$ drops down.

This is shown in Fig. \ref{SN_S2IT_8_20} where the SN rates at
8 and 20 kpc are reported for our Galaxy.
\begin{figure} 
 \centering \includegraphics[scale=0.55,angle=-90]{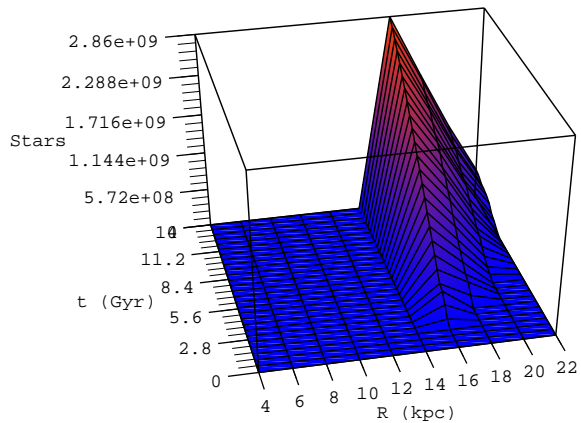}
    \caption{The total number of stars having Earths ($N_{\star life}$) as
      a function of the galactocentric distance and the galactic time
      including the effects of SN explosions described by the case 1)
      prescription for the classical model M31B.   The ($N_{\star life}$) values  are computed within concentric rings, 2 kpc wide.}
		\label{map_M31B_SN1}
\end{figure} 

\begin{figure}
	  \centering   
    \includegraphics[scale=0.35]{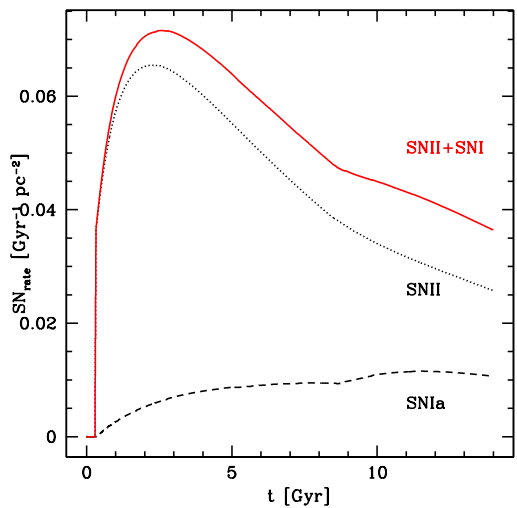} 
    \caption{The Type Ia and II SN rates expressed in pc$^{-2}$
      Gyr$^{-1}$ as a function of the galactic time (Gyr), as
      predicted by the classical model M31B for Andromeda of Spitoni et
      al. (2013) at 8 kpc.}
		\label{SNrate_M31B_8}
\end{figure} 

 In Fig. \ref{S2IT_SN2_t} it is shown that the probability
  $P_{GHZ}$ increases in the outer regions as time goes by, in
  agreement with the previous works of of Lineweaver et al. (2004) and
  Prantzos (2008). In both works it was also found that the peak of
  the maximum probability moves outwards with time. With our chemical
  evolution model such an effect is not present, and the peak is
  always located at the 10 kpc from the Galactic center. This effect
  is probably due to the balance of the destruction by SNe and SFR
  occurring at this distance.
 
 In Fig. \ref{map_S2IT_SN2} we present our results concerning the
 quantity $N_{\star life}$, i.e. total number of stars as a function
 of the Galactic time and the Galactocentric distance for the model
 S2IT in presence of the SN destruction effect (case 2). As found by
 Prantzos (2008) the GHZ, expressed in terms of the total number of
 host stars peaks at galactocentric distances smaller than in the case
 in which are considered the fraction of stars (eq. 3). This is due to
 the fact that in the external regions the number of stars formed at
 any time is smaller than in the inner regions. In fact, the maximum
 numbers of host stars peaks at 8 kpc whereas the maximum fraction of
 stars peaks at 10 kpc (see Fig. \ref{S2IT_SN2_t}).  Our results are
 in perfect agreement with Lineweaver et al. (2004) who identified for
 the Milky Way the GHZ as an annular region between 7 and 9 kpc from
 the Galactic center that widens with time.

\begin{figure*}
	  \centering   
    \includegraphics[scale=0.3]{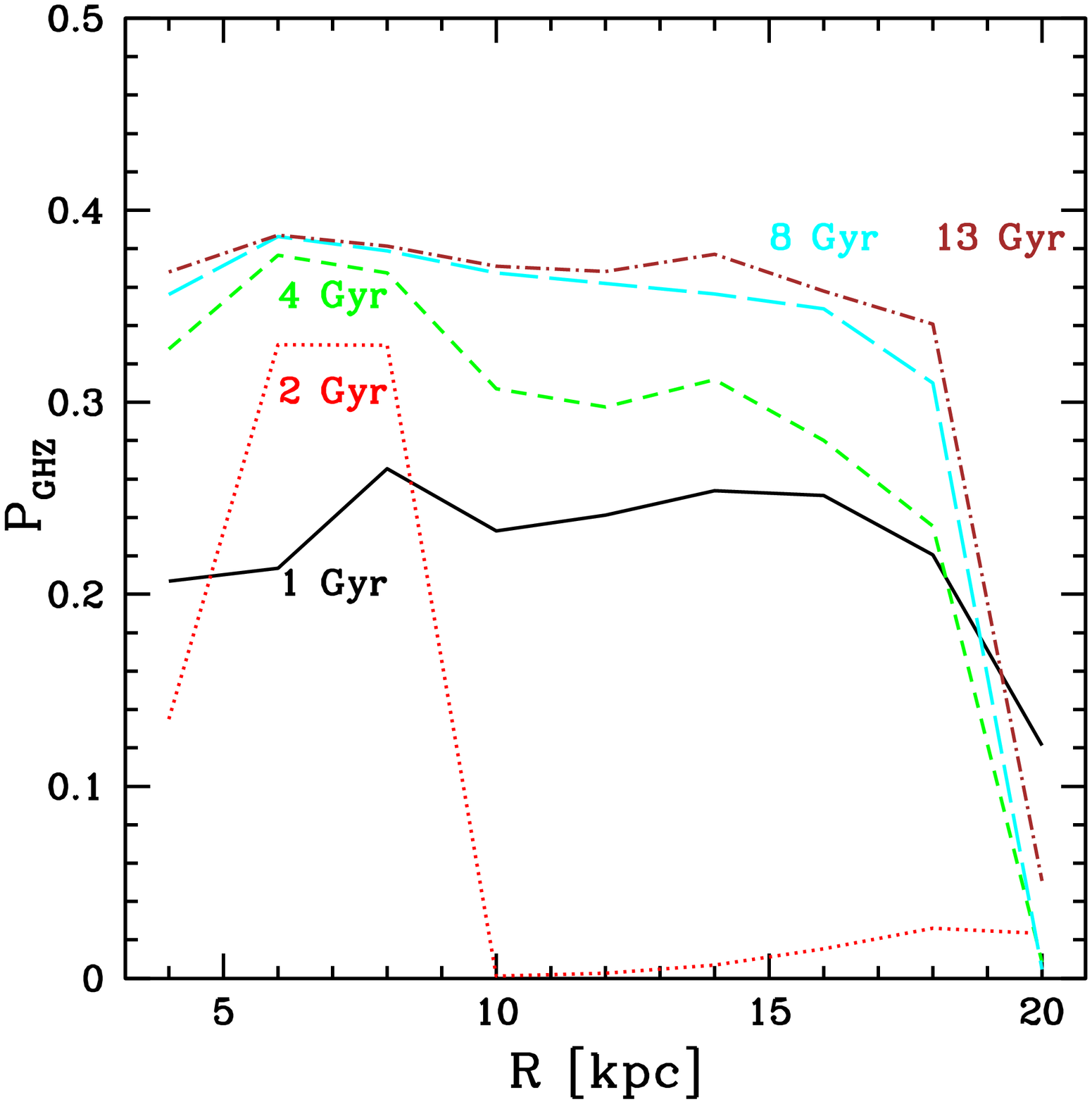} 
  \includegraphics[scale=0.3]{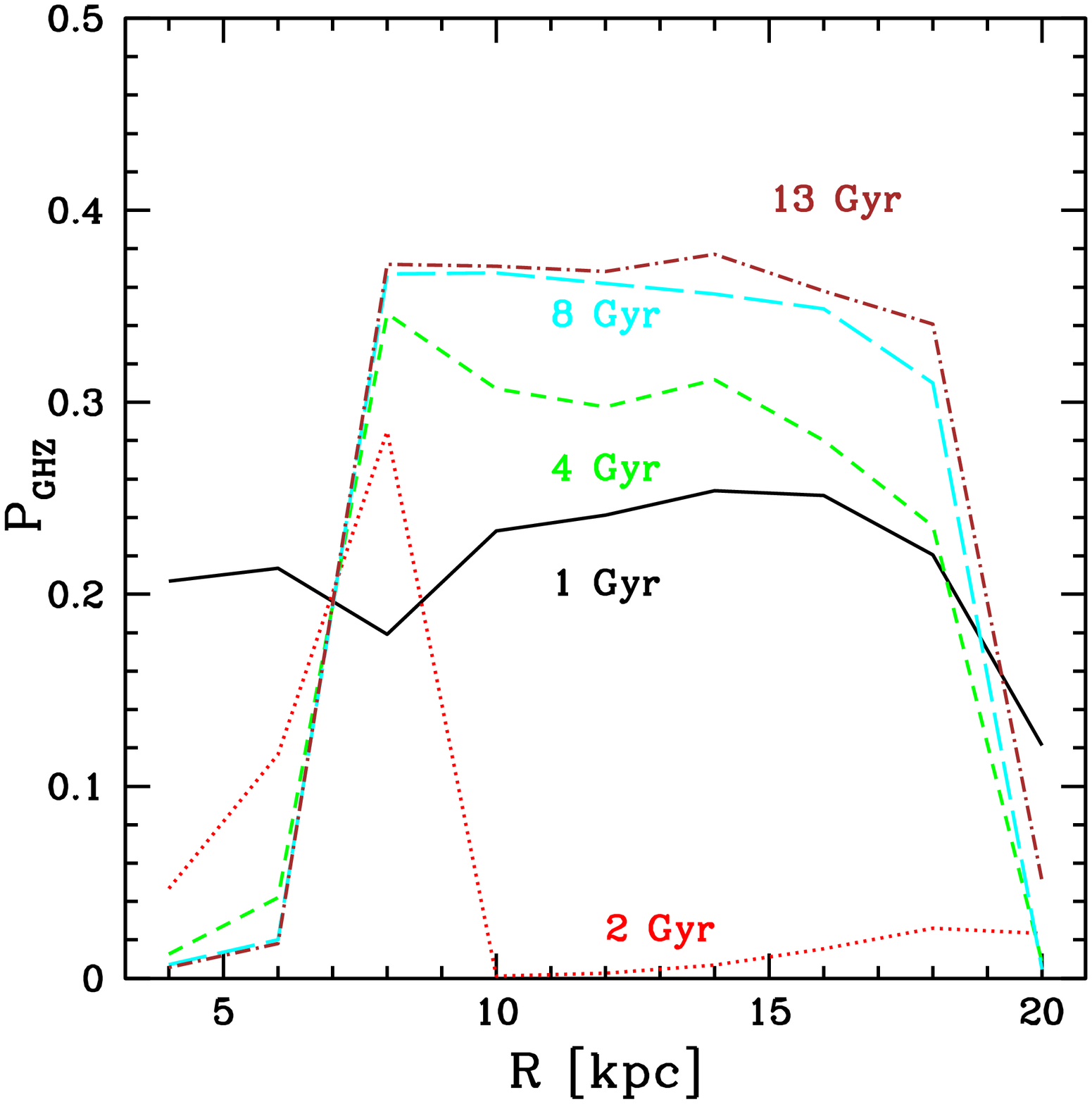} 
    \caption{{\it Left panel}: The probability $P_{GHZ}(R,t)$ as a
      function of the galactocentric distance of having Earths where
      the effects of SN explosions are not considered for the Milky
      Way model RD with radial gas flows at 1, 2, 4, 8, 13 Gyr. {\it
        Right panel}: The probability $P_{GHZ}(R,t)$ as a function of
      the galactocentric distance of having Earths including the
      effects of SN explosions of the case 2) prescription for The
      Milky Way model RD with radial gas flows at 1, 2, 4, 8, 13
      Gyr. }
		\label{RD_NOSN_t}
\end{figure*}

\subsection{M31 model results}

 In Fig. \ref{M31B_NOSN_t} we show the probability
$P_{GHZ}(R,t)$ as a function of the galactocentric distance of having
Earths where the effects of SN explosions are not considered, for the
classical model M31B at 1, 2, 4, 8, 13 Gyr.

 The shape and evolution in time of the $P_{GHZ}(R,t)$  for M31 is similar to the one found by  Prantzos (2008) for the Milky Way.

The similar behavior is due to  the choice of similar
prescriptions for the SFR.  In fact for M31,
as it can be inferred  in Table \ref{TM31}, we have a similar law for
the SFR, in fact our SF efficiency is a function of the galactocentric
distance with the following law: $\nu$= 24/(R[kpc])-1.5.

In Fig. \ref{M31B_NOSN_t} we can see that as time goes by
non-negligible $P_{GHZ}$ values extend to the external parts of
Galaxy. Early on, at 1 Gyr non zero values of $P_{GHZ}$ can be found
just in the inner regions for distances smaller than 10 kpc from the
galactic center. In this plot, it can be visualized the gradual
extension of the fraction of stars with habitable planets up to the
outer regions, during the galactic time evolution.

 The GHZ for the classical model of M31 taking into account  case
1) for the SN destruction effects, is reported in the left panel of
Fig. \ref{M31B_SN1_t}.

 Comparing our results with the ones of Carigi et al. (2013) with the
 same prescriptions for the SN destruction effect (their Fig. 6, first
 upper panels), we find that substantial differences in the inner
 regions (R $\leq$ 14 kpc): at variance with that paper we have a high
 enough SN rate to annihilate the life on formed planets.  In
 Fig. \ref{map_M31B_SN1}, where the total number of stars having
 Earths ($N_{\star life}$) is reported, it is clearly shown that the
 region with no host stars spans all galactocentric distances smaller
 or equal than 14 kpc during the entire galactic evolution. On the
 other hand, the two model are in very good agreement for the external
 regions.

 We have to remind that at variance with our models, the Carigi et
 al. (2013) one is not able to follow the evolution of [Fe/H], because
  they did not consider Type Ia SN explosions.

In the total budget of the SN rate they did not consider the
SN Type Ia. In Fig. \ref{SNrate_M31B_8} we show the contribution of
Type Ia SN rate expressed in Gyr$^{-1}$ pc$^{-2}$ for the M31B model
at 8 kpc. We note that it is not negligible and it must be taken into
account for a correct description of the chemical evolution of the
galaxy.

 It is fair to recall here that our model M31B overestimates the
 present-day SFR.   Therefore, we are aware that in recent time
   there could have been favorable conditions for the growth of the
   life as found in Carigi et al. (2013) work also in regions at
   Galactocentric distances $R \leq 14$. Anyway, as stated by Renda
 (2005) and Yin et al. (2009), the Andromeda galaxy had an overall
 higher star formation efficiency than the Milky Way. Hence, during
 the galactic history the higher SFR had probably led to not favorable
 condition for the life.

The case 2) model is reported in the right panel of Fig.
\ref{M31B_SN1_t}. As expected, the habitable zone region increases in
the inner regions, reaching non-zero values for radius $> 10$ kpc.

 At variance with case 2) for S2IT model, we see that the external
  regions are not affected by the SN destruction. In fact, the SF
  efficiency is lower in the external regions when compared with those
  of the Milky Way. Moreover, the SN rates from the galaxy evolution
  of M31 are always below the value fixed by the case 2): $2
  \times<RSN_{SV}>$, and consequently $P_{GHZ}$ does not change.

\section{Radial flows GHZ results}

We pass now to analyze our results concerning both   M31 and our Galaxy 
in presence of radial flow of gas.

\subsection{The Milky Way  model results in presence of radial flows}

  In the left panel of Fig. \ref{RD_NOSN_t} the RD model results
  without SN destruction are presented.  Although the RD model does
  not include any threshold in the SF, at 20 kpc we note a deep drop
  in the probability $P_{GHZ}(R,t)$, at variance with what we have
  shown for the ``classical'' model S2IT without SN effects
  (Fig. \ref{S2IT_NOSN_t}).

\begin{figure}
  \centering   
  \includegraphics[scale=0.35]{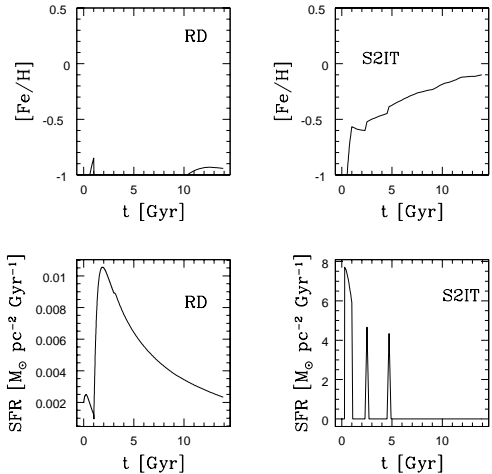} 
  \caption{Left: {\it Upper panel}: [Fe/H] as a function of the
    Galactic time for the model RD at 20 kpc. The lower limit in the
    [Fe/H] is fixed at -1 dex. Below this values the probability of
    forming Earth-like planets ($P_{FE}$) is zero. {\it Lower panel}:
    Star formation rate history for the Galaxy for the model RD at 20
    kpc. Right: the same quantities are plotted for the S2IT model at
    20 kpc.}
		\label{20_window4}
\end{figure}

The explanation can be found in Fig. \ref{vel}: in the outer parts of the Galaxy
the inflow velocities are roughly 2.5 times larger than the M31 ones,
creating a tremendous drop in the SF.   In Fig. \ref{20_window4}
  we report the SFR and [Fe/H] histories at 20 kpc for RD and S2IT
  models respectively.  For the [Fe/H] plot we fixed the lower limit
  at -1 dex, because this is the threshold for the creation of habitable
  planets.

 \begin{figure}
	  \centering   
    \includegraphics[scale=0.35]{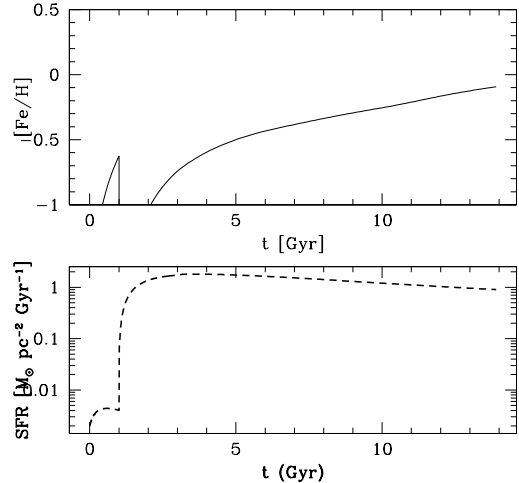} 
    \caption{{\it Upper panel}: [Fe/H] as a function of the Galactic
      time for the model RD at 10 kpc. The lower limit in the [Fe/H]
      is fixed at -1 dex. Below this values the probability of forming
      Earth-like planets ($P_{FE}$) is zero. {\it Lower panel}: Star
      formation rate history for the Galaxy for the model RD at 10 kpc.}
		\label{10kpc_w}
\end{figure}

Concerning the RD model the high inflow velocity has the effect of
removing a not negligible amount of gas from the shell centered at 20
kpc (the outermost shell for the Milky Way model). In the left lower
panel of Fig. \ref{20_window4} we see that the maximum value of the
SFR for the RD model is 0.01 M$_{\odot}$ pc$^{-2}$ Gyr$^{-1}$. It is
important also to recall here that in the RD model for Galactocentric
distances $\geq$ 10 kpc a constant surface mass density of $\sigma_H$
0.01 M$_{\odot}$ pc$^{-2}$ Gyr$^{-1}$ is considered for the halo, at
variance with the model S2IT where it has been fixed at 17 M$_{\odot}$
pc$^{-2}$ Gyr$^{-1}$.

\begin{figure} 
	  \centering   
    \includegraphics[scale=0.6, angle=-90]{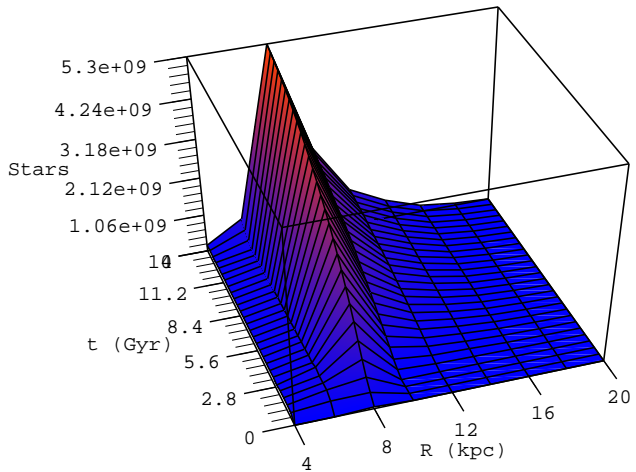} 
   \caption{The total number of stars  having Earths ($N_{\star life}$) as a function of the Galactocentric distance and the Galactic time 
     including the effects of SN explosions described by the case 2)
     prescription for the Milky Way model with gas radial flows (RD
     model).  The ($N_{\star life}$) values  are computed  within concentric rings, 2 kpc wide.}
		\label{map_RD_SN2}
\end{figure}

 The $P_{GHZ}$ value  depends on the product of the [Fe/H] and SFR
 quantities. Because  the RD model at 20 kpc shows [Fe/H] $>$ -1 
 in small ranges of time ($0.6<t<1$ Gyr and $t >
 10.3$ Gyr), we find very small values for $P_{GHZ}$ during the whole
 Galactic history.

 In the right side of Fig. \ref{20_window4} the different behavior of
 the model S2IT at 20 kpc is reported. In this case even if there is a
 threshold in the star formation we have [Fe/H] $>$ -1 dex at all
 times, and the SFR when is not zero is also 800 times higher than in
 the RD model. This is why in Fig. \ref{S2IT_NOSN_t} the model S2IT
 shows higher values of $P_{GHZ}$ at 20 kpc.

 We discuss now the probability $P_{GHZ}(R,t)$ at 2 Gyr in the model
 RD. It drops almost to zero for galactocentric distances $\geq$ 10
 kpc. In Fig. \ref{10kpc_w} we report the SFR and [Fe/H] histories for
 the model RD at 10 kpc.  We recall that in the model RD the halo
 surface density is really small for $R \geq 10$ kpc (0.01 M$_{\odot}$
 pc$^{-2}$ Gyr$^{-1}$).

 We can estimate the $P_{GHZ}$(10 kpc, 2 Gyr) value obtained with
 eq. (\ref{GHZ_NOSN}) using simple approximated analytical
 calculations. For the numerator we note that in the interval of time
 between 0 and 2 Gyr, [Fe/H] $>$ -1 dex only approximately in the
 range 0.5, 1 Gyr, where $P_E$=0.4. In the lower panel of Fig.
 \ref{10kpc_w} it is shown that for this interval the SFR value.  is
 roughly constant and $\simeq 4 \times 10 ^{-4}$ M$_{\odot}$ pc$^{-2}$
 Gyr$^{-1}$.

 Hence, we can
approximate the numerator integral as:

\begin{eqnarray}
\label{}
 \nonumber
N \, \!&=&\! \,\int_0^{2 Gyr} SFR \times P_E \, \, dt'
 \\
  \, \! & \simeq&  \! \, \int_{0.5 Gyr}^{1 Gyr}  4 \times 10 ^{-3}  \times  0.4 \, \, dt' = 0.8 \times 10 ^{-3}.
\end{eqnarray}

The denominator of eq. (\ref{GHZ_NOSN}) is 

\begin{equation}
 D=\int_0^{2 Gyr} SFR \times P_E  \, \,  dt'.
\label{}
\end{equation}

The integral of the SFR is negligible in the first Gyr compared with the
one computed in the interval between 1 and 2 Gyr. A lower limit
estimate of D is given by:

\begin{equation}
 D \simeq \int_{1 Gyr}^{2 Gyr} SFR \times P_E   \, \,  dt' \geq    (1.36) \times 0.5 =0.9, 
\label{}
\end{equation}

where  a linear growth of the SFR from 0 to 1.35 M$_{\odot}$
 pc$^{-2}$ Gyr$^{-1}$ in the interval 1-2 Gyr was considered in reference of the   Fig. \ref{10kpc_w}.

Finally, the approximated   $P_{GHZ}$(10 kpc, 2 Gyr) value is:

 \begin{equation}
\frac{N}{D}\simeq  0.9 \times 10 ^{-3}.
\label{GHZ3}
\end{equation} 

Our numerical result is in perfect agreement with this lower limit
approximation, in fact numerically we obtain $P_{GHZ}$(10 kpc, 2 Gyr)=
1.12609298 $\times 10^{-3}$.

\begin{figure}
	  \centering   
    \includegraphics[scale=0.35]{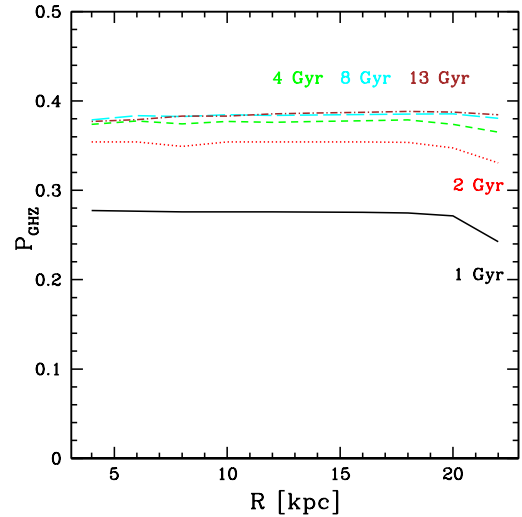} 
    \caption{The probability $P_{GHZ}(R,t)$ as a function of the
      galactocentric distance of having Earths where the effects of SN
      explosions are not considered for the model M31R with radial gas
      flows at 1, 2, 4, 8, 13 Gyr.}
		\label{M31R_NOSN_t}
\end{figure} 

 Here again, as done for the Milky Way ``classical'' model, we report the
 results for the best model in presence of SN destruction which is
 capable to predict the existence of stars hosting Earth-like planets
 in the solar neighborhood.   The SN destruction case 2) gives again  the best results (see Fig. \ref{RD_NOSN_t}). We notice that the external
 regions are not affected from the SN destruction at variance with
 what we have seen above for the ``classical'' model S2IT.

 Comparing the ``classical'' model of Fig. \ref{S2IT_SN2_t} with the
   right panel of Fig. \ref{RD_NOSN_t} we see that for the Milky
   Way the main effect of a radial inflow of gas is to enhance the
   probability $P_{GHZ}$ in outer
   regions when the supernova destruction are also taken into
   account.

 In Fig.  \ref{map_RD_SN2} the quantity $N_{\star life}$ is drawn for
 the RD model. We see that the region with the maximum number of host
 stars centers at 8 kpc, and that this number decreases toward the
 external regions of the Galaxy.  The reason for this is shown in the right
 panel of Fig. \ref{vel}: in the RD model the radial inflow of gas is
 strong enough in the external parts of the Milky Way to lower the
 number of stars formed. Hence, although the probability $P_{GHZ}$ is
 still high in at large Galactocentric distances, the total number of
 the stars formed is smaller during the entire Galactic history compared
 to the internal Galactic regions.  At the present time, at 8 kpc the total
 number of host stars is increased by 38 \% compared to the S2IT
 model results.

\subsection{M31 model results in presence of radial flows}
The last results are related to the GHZ of M31 in presence of radial
flows of gas.  In Fig. \ref{M31R_NOSN_t} we report the probability
$P_{GHZ}(R,t)$ of having Earths without the effects of SN explosions
for the model M31R at 1, 2, 4, 8, 13 Gyr. The main effect of the gas
radial inflow with the velocity pattern of eq. (\ref{linear}), is to
enhance at anytime the probability to find a planet with life around a
star in outer regions of the galaxy compared to the classical result
reported in Fig. \ref{M31B_NOSN_t} also when the destruction
  effect of SN is not taken into account.

 This behavior is due to two main reasons: 1) the M31R model has a
  constant SFR efficiency as a function of  the galactocentric distances, this
  means higher SFR in the external regions; 2) the radial inflow
  velocities are small in the outer part of M31 compared to
  the one used for the Milky Way model RD (see Fig. \ref{vel}),
  therefore the gas removed from the outer shell in the M31R model
  (the one centered at 22 kpc) is very small. In fact, in
  Fig. \ref{M31R_NOSN_t} the drop in the $P_{GHZ}(R,t)$ quantity at 22
  kpc is almost negligible if compared to the one reported in
  Fig. \ref{RD_NOSN_t} for the RD model of the Milky Way, described in
  the Section 6.1.  

For the M31R model with radial flows we will show just the results with
the case 1) with SN destruction.

\begin{figure}
	  \centering   
    \includegraphics[scale=0.35]{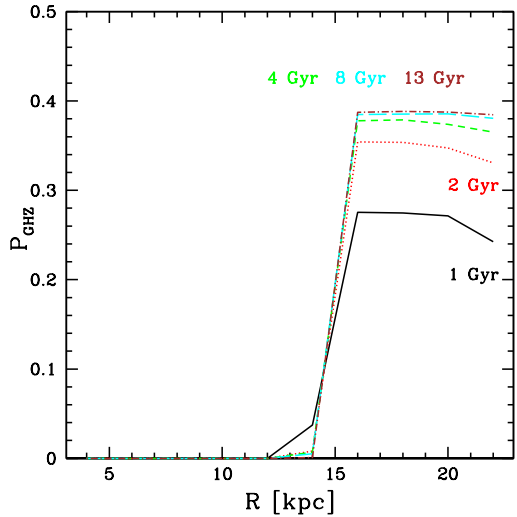} 
    \caption{The probability $P_{GHZ}(R,t)$ as a function of the galactocentric distance of having Earths  including the effects of SN explosions of the case 1) prescription  for the model M31R with radial gas flows at 1, 2, 4, 8, 13 Gyr.}
		\label{M31R_SN1_t}
\end{figure}

\begin{figure} 
	  \centering   
    \includegraphics[scale=0.55, angle=-90]{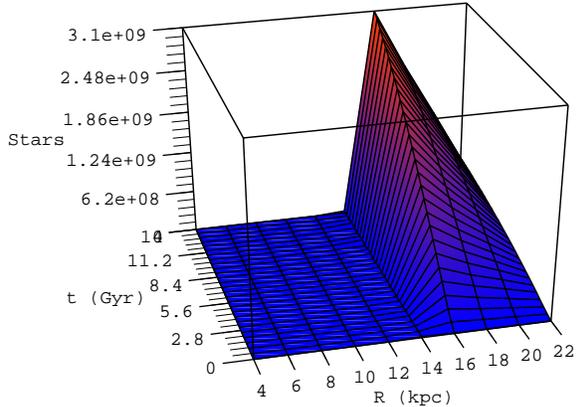} 
    \caption{The total number of stars  having Earths ($N_{\star life}$) as a function of the galactocentric distance and the galactic time 
      including the effects of SN explosions described by the case 1)
      prescription for the model M31R with gas radial flows.  The ($N_{\star life}$) values  are computed within concentric rings, 2 kpc wide.}
		\label{map_M31R_SN1}
\end{figure}

 In Fig. \ref{M31R_SN1_t} the effects of the case 1) SN destruction on
 the M31R model are shown. The external regions for galactocentric
 distances $\geq$ 16 kpc are not affected by SNe, on the other hand
 for radii $<$ 14 kpc due to the higher SNR relative to $<RSN_{SV}>$,
 there are not the condition to create Earth-like planets at during
 whole the galactic time.   In Fig. \ref{map_M31R_SN1} the total
   number of host stars $N_{\star life}$ are shown. Also in this case,
   in the external parts the total number of  stars formed and
   consequently the ones hosting habitable planets are small compared
   to the inner regions. The galactocentric distance with the maximum
   number of host stars is 16 kpc.  Presently, at this distance the
   total number of host stars $N_{\star life}$ is increased by 10 \%
   compared to the M31B model results.

Comparing Fig. \ref{map_M31R_SN1} with Fig.  \ref{map_M31B_SN1} we
note that at variance with the Milky Way results the $N_{\star life}$
value for the M31R model is always higher compared to the M31B
results. This is due to the slower inflow of gas and the constant star
formation efficiency which favor the formation of more stars.
 
For the case 2), we just mention that as expected the GHZ is wider
than the case 1) described above and for radii $<$ 12 kpc, there are
not the condition to create Earth-like planets at during whole the
galactic time.

\section{Conclusions}

In this paper we computed the habitable zones (GHZs) for our Galaxy
and M31 systems, taking into account ``classical'' models and models with
radial gas inflows.  We summarize here our main results presented in
the previous Sections.  Concerning the ``classical'' models we obtained:
\begin{itemize}

\item The Milky Way model which is in agreement with the work of
  Lineweaver et al. (2004) assumes the case 2) for the SN destruction
  (the SN destruction is effective if the SN rate at any time and at
  any radius is higher than two times the average SN rate in the solar
  neighborhood during the last 4.5 Gyr of the Milky Way's life).  With
  this assumption, we find that the Galactic region with the highest
  number of host stars of an Earth-like planet is between 7 and 9 kpc
  with the maximum localized at 8 kpc.
\item For Andromeda, comparing our results with the ones of Carigi et
  al. (2013), with the same prescriptions for the SN destruction
  effects, we find  substantial differences  in the inner
  regions (R $\leq$ 14 kpc). In particular, in this region there is   a high enough SN rate to annihilate
   life on formed planets at variance with Carigi et al. (2013). Nevertheless, we are in agreement for the
  external regions. It is important to stress the most important limit
  of the Carigi et al. (2013) model: Type Ia SN explosions were not
  considered. We have shown instead,  that this quantity is important
  both for  Andromeda and for the Galaxy.

\end{itemize}

 In this work for the first time the effects of radial flows of
  gas were tested on the  GHZ evolution, in the framework of
  chemical evolution models.
\begin{itemize}
\item Concerning the models with radial gas flows both for the Milky Way and
M31 the effect of the gas radial inflows is to enhance the number of
stars hosting a habitable planet with respect to the ``classical'' model
results in the region of maximum probability for this occurrence,
relative to the classical model results.

In more details we found that:

\item At the present time, for the Milky Way if we treat the SN
  destruction effect following the case 2) criteria, the total number
  of host stars as a function of the Galactic time and Galactocentric
  distance tell us that the maximum number of stars is centered at 8
  kpc, and the total number of host stars is increased by 38 \%
  compared to the ``classical'' model results.

\item In M31 the main effect of the gas radial inflow is to enhance at
  anytime the fraction of stars with habitable planets, described by
  the probability $P_{GHZ}$, in outer regions compared to the
  classical model results also for the models without SN
  destruction. This is due to the fact that: i) the M31R model has a
  fixed SFR efficiency throughout all the galactocentric distances,
  this means that in the external regions there is a higher SFR
  compared to the ``classical'' M31B model; ii) the radial inflow
  velocities are smaller in the outer part of the galaxy compared to
  the ones used for the Milky Way model RD, therefore not so much gas
  is removed from the outer shell.  The galactocentric distance with
  the maximum number of host stars is 16 kpc.  Presently, at this
  distance the total number of host stars is increased by 10 \%
  compared to the M31B model results.  These values for the M31R model
  are always higher than the M31B ones.

In spite of the fact that in in the future it will be very unlikely to observe habitable planets in M31,
that could confirm these model results about the GHZ, our aim
here was to test how GHZ models change in accordance with different
types of spiral galaxies, and different chemical evolution
prescriptions.

\end{itemize}

\section*{Acknowledgments}
We thank the 	anonymous  referee  for his suggestions which have
improved the paper. E. Spitoni and F. Matteucci acknowledge financial
support from PRIN MIUR 2010-2011, project “The Chemical and dynamical
Evolution of the Milky Way and Local Group Galaxies”,
prot. 2010LY5N2T.

\end{document}